\documentclass[pdflatex,sn-vancouver,Numbered]{sn-jnl}%
\pdfoutput=1
\usepackage[utf8]{inputenc}
\usepackage[english]{babel}
\usepackage[T1]{fontenc}
\usepackage{amsmath}
\usepackage{amsthm}%
\usepackage{amssymb,amsfonts}
\usepackage{hyperref}
\usepackage{tikz}
\usepackage{lipsum}
\usepackage{pgfplots}
\usepackage{tabularx}
\usepackage{array}

\everymath=\expandafter{\the\everymath\displaystyle}
\IfFileExists{scrextend.sty}{
\usepackage[fontsize=10.000000pt]{scrextend}
}{
\renewcommand{\normalsize}{\fontsize{10.000000}{12.000000}\selectfont}
\normalsize
}
\makeatletter\@ifpackageloaded{underscore}{}{\usepackage[strings]{underscore}}\makeatother
\newtheorem{theorem}{Theorem}
\theoremstyle{definition}
\newtheorem{remark}{Remark}

\newcommand{\1}{\mathbf{1}}

\newtheorem{statement}[theorem]{Statement}%
\pgfplotsset{compat=1.17} 
\begin{document}

\title{A quantum-classical hybrid branch \& bound algorithm}

\author*[1]{\fnm{Andr\'as} \sur{Cz\'egel}}

\author[2]{\fnm{D\'avid} \sur{Sipos}} 

\author[1]{\fnm{Bogl\'arka} \sur{G.-T\'oth}}

\affil*[1]{\orgdiv{Department of Computational Optimization}, \orgname{University of Szeged}, \orgaddress{\street{\'Arp\'ad t\'er 2.}, \city{Szeged}, \postcode{6720}, \country{Hungary}}}

\affil*[2]{\orgname{Hamburg University of Technology}, \orgdiv{Institute for Algorithms and Complexity}, \orgaddress{\city{Hamburg}, \postcode{21073}, \country{Germany}}}

\abstract{We propose a complete quantum-classical hybrid branch-and-bound algorithm (QCBB) to solve binary linear programs with equality constraints. That includes bound calculation, convergence metrics, and optimality guarantee to the quantum optimization based algorithm, which makes our method directly comparable to classical methods. Key aspects of the proposed algorithm are (i) encapsulation of the quantum optimization method, (ii) utilization of noisy samples for problem reduction, (iii) classical approximation based bound calculation, (iv) branch and bound traits like gap-based stopping criterion and monotonic increase in solution quality, (v) integrated composition of many different solutions that can be improved individually. We show numerical results on set partitioning problem instances and provide many details about the characteristics of the different steps of the algorithm.}

\maketitle

\section{Introduction}

Quantum optimization is one of the most rapidly advancing fields of quantum technology. It has been in the headlines because of the many promises over classical algorithms on certain problems. A comprehensive article recently investigated the different approaches and their challenges \cite{Abbas2024}.
When we look for solutions to Integer Linear Programming (ILP) problems, most quantum optimization algorithms, like Quantum Approximate Optimization Algorithm (QAOA) family \cite{Farhi2014, Hadfield2019, Chalupnik2022}, the variational Quantum Imaginary Time Evolution (varQITE) algorithm \cite{Motta2019, Soley2021, Gacon2024} and the Quantum Iterative Power Algorithms (QIPA) \cite{Kyaw2023} are designed to tackle the whole problem at once, even though, utilizing different principles.  Classical optimization solvers are usually composed of many different algorithms \cite{Andersen1995, Achterberg2020, KlnKarzan2009, Achterberg2005, Berthold2025, Turner2023, Zeng2025} to solve carefully constructed subproblems. Our aim is to adapt this approach to quantum optimization solutions as well. 

To provide context for our hybrid approach, we take a brief look at the literature, investigating ideas related to decomposition-based and sequential hybrid optimization.

There has been a recursive decomposition scheme introduced to QAOA by Bravyi et al.\ in \cite{Bravyi2020, Bravyi2022}. In their algorithm, RQAOA, correlations are computed in the physical system, and then the system size is reduced, with the aim of breaking symmetries, leading to improvement over the original QAOA algorithm. A new algorithm family, Quantum-Informed Recursive Optimization Algorithms (QIRO) \cite{Fingar2024}, goes further along this path: they enhanced decomposition steps by classical, problem-dependent rules and backtracking. Computing correlations can suffer from noise and algorithmic error of the QAOA, leading to wrong decisions when reducing the systems. To some extent, the backtracking in QIRO resolves this issue, however, it does not build a complete tree during recursive simplifications, thus it does not guarantee finding optimal solutions. There is also a promising recent work on how deep learning can enhance the performance of variational quantum optimizers \cite{Zhang2025}, although the algorithm does not guarantee optimality either.

While these algorithms apply a recursive decomposition, another idea of enhancing variational quantum optimization algorithms is by classically computing warmstart values to the variational parameters \cite{Egger2021}, which also leads to significant improvements over the original variational quantum algorithm.

The other side of finding new hybrid optimization solutions is by classical techniques from Operations Research. These are powerful tools, even though they do not enhance the quantum optimization routine itself, but utilize it better. One such approach is to use Benders decomposition to split the mixed ILP problem into a continuous linear program and a quadratic unconstrained binary problem designed to solve on quantum computers \cite{Zhao2022}. There were also studies towards a quantum branch and bound algorithm \cite{Montanaro2020, Chakrabarti2022} by using quantum tree search \cite{Jarret2018} and estimating the tree size \cite{Ambainis2017}, resulting in a near-quadratic speedup on the tree search. This algorithm, despite belonging to the same family of algorithms, uses a vastly different approach from our branch and bound construction in each of its main steps. 

We have also proposed a conflict-driven constraint generation algorithm for binary linear programs \cite{Czegel2025}. In that idea, the quantum algorithm is used to solve an unconstrained relaxation of the original problem, and then iteratively adding back constraints until the quantum algorithm successfully finds high-quality feasible solutions.

In this work, we take a step further along this path of wrapping an arbitrary variational quantum optimization algorithm. We propose a conflict-driven branch and bound method that has a decomposition structure with bounds and guarantees. We solve each subproblem with a quantum device that provides solution candidates by sampling its solutions state. From those samples, we derive a quantity for each variable, which tells us how much complication it holds, i.e.\ how many conflicting constraints it is involved in. We call them conflict values, and we use them to select branching variables. We also use constraint propagation to further reduce the problem size at each branching step. By combining these techniques, we aim to reduce the complexity and the size of the problem at each step, driving the quantum algorithm towards better quality solutions. We also designed a fast classical bounding method that enables us to cut the branch and bound tree, and more importantly, to prove optimality.

With this work, we would like to provide a ground for further advancements. Our work does not aim to top classical solutions on industrially relevant problems, but rather be a proof of concept. We provide a framework that can be highly improved by optimizing its parts for the respective problem, software, and hardware requirements.

Our paper is structured as follows: first, we introduce the problem in Section \ref{sec:problem}, then describe the QCBB algorithm in Section \ref{sec:alg_full}. After that, we provide numerical results in Section \ref{sec:results}. Then, we discuss how the algorithm behaves, discuss improvement opportunities, and provide more contextual information in Section \ref{sec:disc}.

\section{Problem}
\label{sec:problem}

Our initial master problem \eqref{orig} is to find a solution to a binary linear problem, 

\begin{equation}\label{orig}
    \min~ \left\{ c^Tx ~\bigm|~ Ax = b, \ x\in\left\{0,1\right\}^n \right\}, \tag{MP}
\end{equation}
where $c\in\mathbb{R}^n, A\in\mathbb{R}^{m \times n}, b\in\mathbb{R}^m$. This can be generalized to integer variables \cite{Karp1972, Lucas2014}. In the following sections, we propose an algorithm to solve \eqref{orig}.   

\subsection{Master Hamiltonian}\label{sec:master-ham}

We transform \eqref{orig} into an Ising Hamiltonian. For this transformation, we create a Quadratic Unconstrained Binary Optimization (QUBO) form of it, and then transform variables from the $\{0,1\}$ domain to the $\{-1, 1\}$ domain. The transformation is derived in Appendix~\ref{app:BLPtoH}. The Ising Hamiltonian, which we name as Master Hamiltonian \eqref{eq:MH} for future reference, is the following:

\begin{equation}
\label{eq:MH}
    H(\sigma) = - \sum_{i<j}J_{ij}\sigma_i \sigma_{j} - \mu \sum_i h_i \sigma_i, \tag{MH}
\end{equation}
where
\begin{equation}
    J = -\frac{1}{4}MA^TA
\end{equation}
and
\begin{equation}
    h = c^T - 2 Mb^T A + M\mathbf{1}^T (A^T A)
\end{equation}
with
\begin{equation}
    \mu = -\frac{1}{2},
\end{equation}
where big $M$ is a sufficiently large constant, and by putting aside a constant term 
\begin{equation}
    C = \frac{1}{4} M \mathbf{1}^T (A^T A) \mathbf{1} + \frac{1}{2} \mathbf{1}^T c - M b^T A \mathbf{1}
    + M b^T b \label{eq:C}
\end{equation}
from the objective function. One can find exact calculations for big $M$ in \cite{Czegel2025} and
in Appendix \ref{app:bigM}.

\section{Quantum-Classical Branch \& Bound}
\label{sec:alg_full}

Our proposal is an algorithm that uses quantum computing within a branch-and-bound framework. The branch-and-bound is a general optimization framework that searches for the best solution through recursive decomposition of the problem into smaller subproblems (branching). For each subproblem, bounds on the objective value are computed to estimate its potential. Subproblems whose bounds indicate no improvement over the current best solution are discarded (pruned). This recursive branching, bounding, and pruning process continues until all promising regions are explored, thus the global optimum is found without the complete enumeration of all possible solutions.

The algorithm builds a decomposition tree. Note that this is a recursive definition for building an evaluation tree, where in this case the root is the master problem, and subproblems get simpler and smaller along the path to leaf nodes. From now on, let us denote the elements of the problem of the node under evaluation by $\hat{H}, \hat{A}, \hat{b}, \hat{c}, \hat{x}, \hat{\sigma}$ having $\hat{n}$ variables, corresponding to \eqref{eq:MH}-\eqref{eq:C} for the reduced problem \[\min~ \left\{ \hat c^T\hat x ~\bigm|~ \hat A\hat x = \hat b, \ \hat x\in\left\{0,1\right\}^{\hat n} \right\}\tag{RP}\label{eq:RP}.\] We denote child node subproblem elements by a tilde on the respective element.  At any point of time of the execution, the feasible set of the subproblems in the frontier of the tree contains the optimal solution of the problem at the root, the master problem. 

In general, the definition of a branch-and-bound framework requires the specification of the following rules: branching, node selection, bounding, pruning, and termination. Quantum computation is involved in the branching and bounding rules; therefore, we will focus on describing them first.

\subsection{Variational Quantum Solution}

At this point, \eqref{eq:MH} encodes our feasible solutions in its lower energy states. Now we look for a suitable VQA to search for these states. Here, we chose the QAOA for simplicity and clarity, but any VQA suffices.

Our aim is to find the state preparation routine $U$ such that $U|0\rangle^{\otimes n} = |\psi^*\rangle$, where $|\psi^*\rangle$ corresponds to the ground state of the Hamiltonian ideally. In reality, $|\psi^*\rangle$ is rather a superposition of states that we hope to belong to low energy levels, i.e.\ encode feasible solutions to \eqref{orig}. Upon measurement, we get a sample from a probability distribution that the measurement represents with respect to the state.

We can use this behavior to sample from the result state $q$ times to look for good solutions, and also for other information that we can utilize. We denote the number of different samples by $s$ and their respective counts in the samples by $\omega=(\omega_1,\ldots,\omega_s)$ where $\sum_{i=1}^s \omega_i = q$. 

Let us denote the gathered different samples by $X$, where sample $\ell$ is the $\ell$th column of $X$, i.e. where $X_{i\ell}$ is $x_i$ of sample $\ell$, for $\ell=1,\ldots, s, \ i=1,\ldots,n$.

From the samples, the incumbent can be updated. That is, if there is an improving solution among the samples, i.e. the value of it is less than the previous best, that solution becomes the incumbent.

\subsection{Branching}
\label{sec:branching}
Branching means that the selected subproblem is decomposed into smaller problems. Generally, it can be done by selecting a branching variable and separating the problem by restricting its bounds. In order to choose the variable to branch at, we use the gathered information from the samples. 

\subsubsection{Conflict values}

In this step, we derive conflict values for variables. This tells about a variable, that how much constraint violations it has been part of. 

First, we define violation matrix $V \in \left\{0,1\right\}^{m \times s}$, where 

\begin{equation}
    V_{j\ell} = \begin{cases}
        1, \quad\text{when } \sum_i \hat A_{ji}X_{i\ell} - \hat b_j \neq 0 \\ 
        0, \quad\text{otherwise.}
    \end{cases} 
\end{equation}
Note that $V$ does not represent how much these constraints were violated. Since we don't deal with numeric issues here explicitly, the aim is to reduce the number of overall violations. In other words, we don't look for extreme outliers, but for controversial variables.

We can obtain violation scores as dual-like quantities for each constraint $j$ by 
\begin{equation}
    \nu_{j} = \frac1{q}\sum_{\ell=1}^s V_{j\ell}\omega_{\ell} \qquad j=1,\ldots,m.
\end{equation}
This gives a normalized vector $\nu \in [0,1]^m$, where each $\nu_j$ tells us what proportion of the samples violated constraint $j$ of \eqref{orig}. Using $\nu$, we can derive a conflict value for each variable $x_i$ that tells us how much it contributes to constraint violations. For that, let $P$ denote the correspondence matrix for variables in constraints as
\begin{equation}
    P_{ji} = \begin{cases}
        1, \text{ when } \hat A_{ji} \neq 0 \\ 
        0, \text{ otherwise.}
    \end{cases} 
\end{equation}
Then, the conflict vector $\gamma \in \mathbb{R}^{\hat n}$ is computed as 
\begin{equation}
    \gamma = \nu P. 
\end{equation}
In other words, $\gamma_i$ measures for each variable $x_i$, how much it contributes to constraint violations.

Note that these conflict values in general do not depend on the actual objective coefficient of the variable they belong to. However, we can use the information of the variable values later for branch selection. 

\subsubsection{Variable selection}

We select the most conflicting variable to branch on, i.e.\
\begin{equation}
    k = \arg \max_{i} \gamma_i. \label{eq:varsel}
\end{equation}

Restricting variables on their bounds on binary programs also means setting their value exactly. 

\subsubsection{Subproblem construction}
\label{sec:subproblem}

The variable selected by \eqref{eq:varsel} can be removed then from the current problem as if we set $\hat{x}_k = 0$ or $\hat{x}_k = 1$,
\begin{equation}
    \Tilde{b}_j = \hat{b}_j - \hat{A}_{jk}\hat{x}_k \qquad \forall j \label{eq:btilde}
\end{equation}

To update $\hat{A}$ and $\hat{c}$ and remove $\hat{x}_k$, we define the column-elimination matrix $D$ as an $\hat{n} \times \hat{n}-1$ matrix, which is obtained by removing column $k$ from the identity matrix. Now, define the new coefficients of the new subproblem using
\begin{equation}
\label{eq:branchcoeff}
    \Tilde{A} = \hat{A}D, \qquad \Tilde{c} = D^T\hat{c}, \qquad \Tilde{x} = D^T\hat{x}
\end{equation}
that removes variable $\hat{x}_k$ from \eqref{eq:RP} and adjusts the problem structure to its fixed value. 

Consequently, the Reduced Hamiltonian \eqref{eq:RMH} for the new subproblem is
\begin{equation}
\label{eq:RMH}
    \Tilde{H}(\Tilde{\sigma}) = - \sum_{i<j}\Tilde{J}_{ij}\Tilde{\sigma}_i \Tilde{\sigma}_j - \mu \sum_j \Tilde{h}_i \Tilde{\sigma}_i, \tag{RH}
\end{equation}
where
\begin{equation}
    \Tilde{J} = -\frac{1}{4}M\Tilde{A}^T\Tilde{A}
\end{equation}
and
\begin{equation}
    \Tilde{h} = \Tilde{c}^T - 2 M\Tilde{b}^T \Tilde{A} + M\mathbf{1}^T (\Tilde{A}^T \Tilde{A})
\end{equation}
with $\mu = -\frac{1}{2}$. Note that $\Tilde\sigma=D^T\hat{\sigma}$, eliminating the $k$th element of $\hat{\sigma}$. The Hamiltonian is also of size $\hat{n}-1 \times \hat{n}-1$, where $\hat{n}$ denotes the size of the recently evaluated node's problem. While constructing $\Tilde{H}(\Tilde{\sigma})$, we get an additional constant term that we put aside during the quantum optimization step. This constant, however, is needed to obtain the solution to \eqref{orig}. It is composed from the constant part of the transformation to the Ising Hamiltonian,
\begin{equation}
    \Tilde{C}_\text{transform} = \frac{1}{4} M \mathbf{1}^T (\Tilde{A}^T \Tilde{A}) \mathbf{1} + \frac{1}{2} \mathbf{1}^T \Tilde{c} - M \Tilde{b}^T \Tilde{A} \mathbf{1}
    + M \Tilde{b}^T \Tilde{b}, 
\end{equation}
and from the part that comes from the objective coefficient of the eliminated variable,
\begin{equation}
    \Tilde{C}_\text{objective} = \hat C_\text{objective} + \hat{c}_k\hat{x}_k,
\end{equation}
thus the constant term is
\begin{equation}
    \Tilde{C} = \Tilde{C}_\text{transform} + \Tilde{C}_\text{objective}.
    \label{eq:Ctilde}
\end{equation}

\subsubsection{Variable bound propagation}

In binary problems, branching on a variable is not just a restriction, but an exact assignment. By assigning 0 or 1 to a variable, there might be other enforced reductions in the problem. These reductions come from dependence on the assigned variable in other constraints. On the other hand, either the original or the propagated reductions could discover infeasibility, which allows us to prune the whole sub-tree. This direction of reduction is well described in \cite{Fingar2024} and used as propagation logic inside SAT-solvers that are based on the Davis–Putnam–Logemann–Loveland (DPLL) algorithm \cite{Davis1960, Davis1962}.

The propagation of some variables means that the problem becomes even smaller. These variables are treated in the same way as the branching variable, which means that we apply the formulas \eqref{eq:btilde}-\eqref{eq:Ctilde} with all fixed variables, where $D$ becomes a matrix having all columns with fixed variables removed.

\begin{figure*}[ht]
  \centering
  \includegraphics[width=1.0\textwidth]{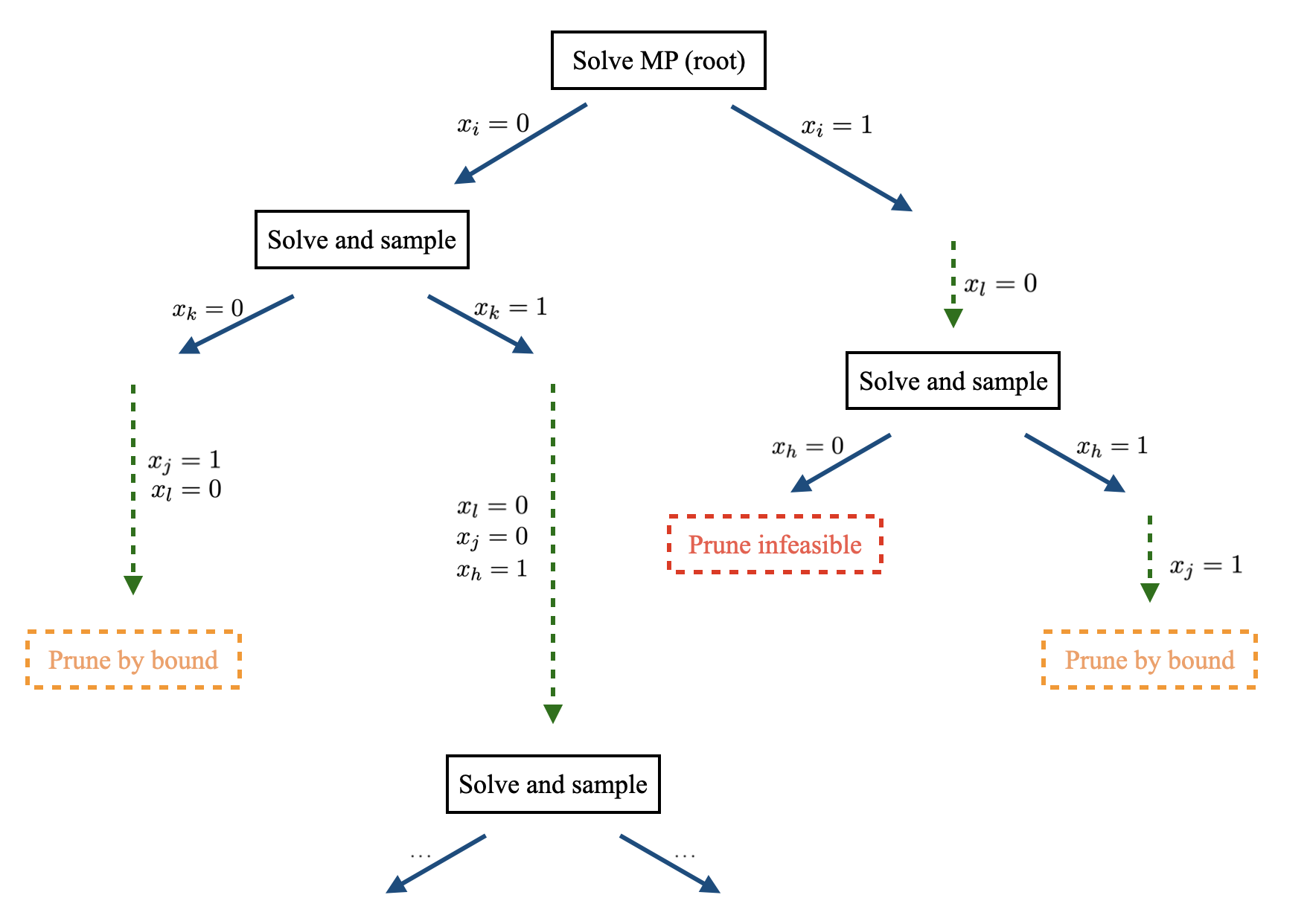}
  \caption{An example for the tree of the branching algorithm. First, the Master Problem is solved. Then, we branch by assigning values to one variable (blue arrows). After that, we try to further reduce the problem by constraint propagation steps, which also assign values to variables (green dashed arrows). 
  In each node, we calculate a bound, and if possible, prune based on that (orange node). We also prune infeasible subproblems (red nodes).}
  \label{fig:tree}
\end{figure*}

\begin{figure}[ht]
  \centering
  \includegraphics[width=0.5\textwidth]{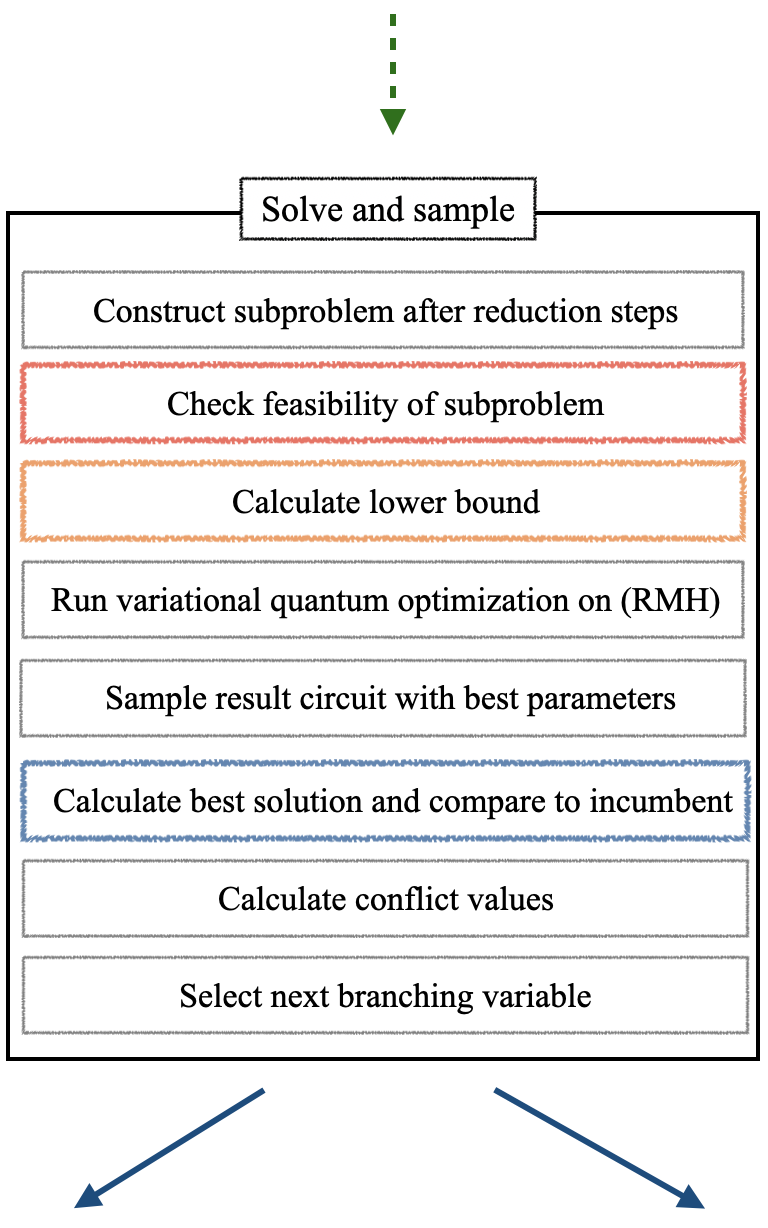}
  \caption{Steps of a single node evaluation in the tree. The boxes with thick red, orange borders note the prune criteria: checking infeasibility and bounding. The blue box indicates the possible update on the incumbent value.}
  \label{fig:solve}
\end{figure}

The complete subproblem creation and propagation creates a tree structure, shown in Figure \ref{fig:tree}, where the process inside a node evaluation is on Figure \ref{fig:solve}. 

\subsection{Node selection}

In this initial work, we choose in each iteration the next node to be evaluated by the lowest lower bound. This is a simple, standard strategy that has some good aspects to it. It gives the greatest opportunity to reduce the gap between the incumbent solution value and the global lower bound. This can be viewed as a greedy selection rule, which minimizes the number of evaluated nodes, trying to close the gap with as few steps as possible. This comes in handy as node evaluation is the most expensive step in our algorithm.

\subsection{Bounding}
\label{sec:bound}

Bounding is an important step for reducing the search space and allowing effective cuts to the tree. Lacking an LP-relaxation and thus a relaxed bound to \eqref{eq:RP} poses a major challenge in getting a meaningful lower bound. Here, we propose an efficient approximation-based lower bound calculation to obtain good lower bounds at each node. The strategy itself is simple and bounds the reduced problem at the given node. Since our branching decisions are variable assignments that lead to problem reductions, we can consider a bound on the reduced objective \eqref{eq:RMH}, corrected with the constant terms to obtain a valid lower bound to the original problem. The bounding has the following three steps: (i) transform \eqref{eq:RMH} to a Maximum Cut problem, (ii) obtain an approximate solution, $z_{GW}$, to the Maximum Cut by the Goemans-Williamson algorithm \cite{Goemans1995}, and (iii) calculate a bound based on the guaranteed approximation ratio.

The lower bound is obtained by
\begin{equation}
    LB_{GW}(\hat H(\sigma)) = -\frac{2}{\alpha} z_{GW} + \left(\frac{2}{\alpha}-2\right) W^- + W,
\end{equation}
where $\alpha = 0.87856$, which is the average worst-case approximation ratio of the Goemans-Williamson algorithm, 

\begin{equation}
    W^- = \sum_{i < j, J_{ij} < 0} \hat J_{ij} + \mu\sum_{i: h_i < 0} \hat h_i, \qquad \text{ and} \qquad
    W = \sum_{i < j} \hat J_{ij} + \mu\sum_i \hat h_i,
\end{equation}
i.e. $W^-$ is the sum of all negative coefficients in $\hat H(\sigma)$ and $W$ is the sum of all coefficients in $\hat H(\sigma)$. If the bound is looser than the bound from the parent node, we take the bound from the parent node. Thus, for any child node problem defined by $\tilde H$ and parent node problem $\hat H$, omitting the variables from the notation for convenience,
\begin{equation}
\label{eq:LB}
    LB(\tilde H) = \max \left\{LB_{GW}(\hat H) , LB(\hat H) \right\}.
\end{equation}

Since this only bounds the Hamiltonian's ground state energy, we need to add back the constant term \eqref{eq:Ctilde} to get a lower bound to the node under evaluation. 
We discuss the calculations and quality guarantees in Appendix \ref{app:bound}. Adding back all the fixed variable values, we end up at a local lower bound for the sub-tree, rooted at the evaluated node. The lowest of these bounds in the frontier of the tree is the global lower bound, due to the disjunctive structure of the subproblems. Note that when we reach a subproblem with no free variables, $LB_{GW}(\hat{H}(\sigma)) = \hat C_\text{objective}=\sum_{k \in \{1..n\}} c_kx_k$, which is then equal to the actual value of the objective.

\subsection{Pruning}

The branch-and-bound framework uses the bound calculation to prune nodes and their sub-tree. If the lower bound of the subproblem at a given node is greater than the incumbent that we got from the variational quantum algorithm, we can prune the tree at the node, because the subproblem can not provide any improvement on the incumbent.

\subsubsection{Infeasibility pruning}

After the branching assignment and possible further assignments from bound propagation, we have more information about the problem that we can use. The set of assignments might reveal contradictions in the constraints. If we find such a contradiction, the subproblem is infeasible, and we can prune the branch.

The individual constraint contradiction itself is sufficient to prove infeasibility, but its absence does not prove feasibility. The subproblem constraints could each be satisfied on their own, while these assignments contradict each other. 

With our current bounding strategy, we can define one more feasibility-based pruning rule, because if $LB_{GW}(\hat H) + \hat C \geq M$, then the subproblem is infeasible. We provide details to this in Appendix \ref{app:infeas}.

Checking and proving feasibility of the subproblem itself is NP-complete \cite{Karp1972}. Despite the lack of a guarantee of feasibility, this is a good condition for pruning.

\subsection{Stopping criteria}

Some quantum algorithms, like the QAOA family \cite{Farhi2014, Hadfield2019, Chalupnik2022} and other time evolution based methods \cite{Motta2019, Soley2021, Gacon2024} offer rigorous error measures and guarantees, however these are hard to realize on current quantum hardware \cite{Bharti2022}. While there are many ways to improve the performance and precision of the quantum optimization routine \cite{Chandarana2022, Mundada2023, Cai2023, Zhu2024}, mathematical guarantees remain difficult to achieve in a practical setting.

The flipside is, hybrid quantum optimization algorithms themselves aim to find optimal solutions to the discrete optimization problem, and not the relaxation. This provides a highly uncertain advantage: they might find good quality, or even optimal solutions early. However, they could, in practice, provide infeasible solutions in all the samples of the output state. In this algorithm, we work on these nodes as well, and continue until the point where where bounds themselves provide proof of infeasibility or non-optimality. At that point, the quantum algorithm can either produce feasible solutions or we can prove the infeasibility of the subproblem itself.

Coming from the scheme of branch and bound, there are some criteria based on which one could stop the run, for example, (i) number of evaluated nodes, (ii) runtime, (iii) proven optimality, and (iv) duality gap target value. The first two are straightforward, measuring the number of evaluated nodes and runtime, and stop upon we reach the limit. Then, return the best solution found so far.

\subsubsection{Proven optimality}

Traditionally, quantum algorithms don't have lower bounds, nor strict convergence guarantees on which they can be stopped. We aim to change that. We obtain a global lower bound, described in Section \ref{sec:bound}. When the global lower bound meets the current upper bound, the incumbent solution, we have a proof of optimality.

\subsubsection{Target gap value}

Of course, reaching optimality can take exponential time, and there are bound-based criteria. One of them is to calculate the ratio between the upper bound and lower bound, defining a gap value. This gap is monotonically decreasing due to the structure of branch and bound. While reaching a given gap might also take exponential time, in reality, reaching a reasonable gap is usually a faster and more viable solution than going for proven optimality. With that said, stopping early still holds valuable information: how far are we, in the worst case, from an optimal solution.

\section{Results}
\label{sec:results}

We applied our method to many different instances of the Set Partitioning Problem (SPP). The problem has practical applications, for example, in scheduling airline crews~\cite{Sipos2025}. 

\subsection{Set Partitioning Problem}
In this problem, given a collection $\mathcal{S}=\{s_1,\dots,s_n\}$ of subsets of a set $R = \{r_1, \dots, r_m\}$ and associated costs $C=\{c_1,\dots,c_n\}$, we are looking for a subset $\mathcal{S}^*\subset \mathcal{S}$ which covers exactly once every element of $R$ at minimum total cost.

By associating every subset $s_i$ with a variable $x_i$, and defining $a_{ij}=1$ if $r_j\in s_i$, 0 otherwise, the problem can be modeled as the following binary linear program,

\begin{align}
    \min& \quad \sum_{i=1}^{n}c_ix_i&\\
    \textrm{s.t.}& \quad \sum_{i=1}^{n} a_{ij} x_i=1& \quad \forall j\in\{1,\dots,m\}\\
    &\quad x_i \in \{0,1\} &\forall i \in \{1,\dots,n\} 
\end{align} .

We transform this formulation as described in Section \ref{sec:master-ham}.

\subsection{System integration}
We utilized the benchmark suite developed in previous work \cite{Sipos2025} to test our algorithm. The solve and sample, and branching steps, illustrated in Figure \ref{fig:tree} and Figure \ref{fig:solve}, are implemented as a plugin in the benchmark suite, which can leverage already existing integrations of quantum computing providers, optimizers, and QAOA ans\"atze while also being usable on its own as a Python module.
Execution of the algorithm on different problems is conducted through what is referred to as an \texttt{Experiment} in the suite, which allows the creation and reuse of configuration files to produce and extract insights about the algorithm's behavior, which we expand on in the following section.

\subsection{Experimental results}

Keeping in mind constrained computational resources, we tested the algorithm on randomly generated instances of SPP.
For every instance, at least a single feasible solution is guaranteed and no instance contains the complete set, i.e. $s_i \subset R \ \forall i$. All problem instances have $n=15$ variables (subsets) and a varied number of constraints ( $m$ elements), resulting in a quantum circuit with $15$ qubits.
In our testing, we used the QAOA algorithm~\cite{Farhi2014}, at the depth of $p=3$, for simplicity with no platform-specific circuit, or hardware optimizations, as we focus on the classical framework in this work. For optimizing the expectation value of the quantum state, we use the COBYLA \cite{Powell1994} algorithm.

\begin{figure}[h]
    \centering
    \includegraphics{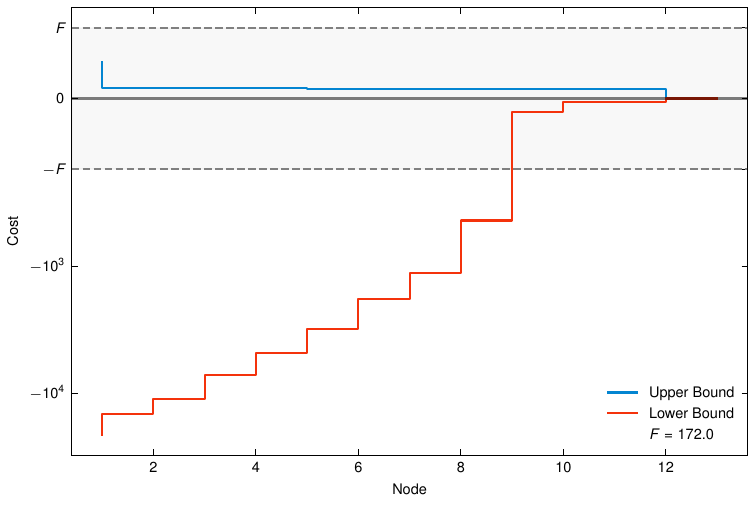}
    \caption{Convergence of bounds. The blue line shows the upper bound (cost of incumbent solution) as a function of the number of nodes evaluated. The red line shows the lower bound as a function of the number of nodes evaluated. A solid black line marks the optimum, aligned to $0$, with dashed black lines representing the cost $F$ of the worst feasible solution, in this case $172$. Between the dashed lines, the y-axis is scaled linearly, while outside the dashed line, a logarithmic scale is used.}
    \label{fig:gap}
\end{figure}

For measuring branch and bound algorithms, and in general, classical mixed-integer linear programming (MILP) solvers, there is a concept called the primal-dual integral \cite{Berthold2013}. In short, it measures how the best solution (primal) and the best lower bound (dual) change together over time. Since both converge to an optimal solution value, the area between them measures the quality of the solution process of a solver (the area between the red and blue lines in Figure \ref{fig:gap}). We do not solve the LP-relaxation of the binary program, but, as described in Section \ref{sec:bound}, we obtain a bound based on the equivalent Maximum Cut formulation of the problem.
So in the traditional sense, we do not calculate dual problems, but a monotonic improving lower bound, which qualifies the algorithm to be measured by this metric. 

In Figure \ref{fig:gap}, we show on a particular problem instance how the upper and lower bounds evolve as nodes get evaluated, with further instances shown in Appendix \ref{app:results}. Bound values are adjusted such that the optimum is at $0$, marked by a solid black line. As a consequence of the penalty terms introduced in the QUBO formulation, bound values associated with infeasible solutions are quite large, as such, we opted to use a logarithmic scale a certain distance away from the optimum, marked by dashed black lines. This distance, $F$, is given by the highest-cost feasible solution, and within its distance from the optimum, the scale is linear.
The blue line with x-markers represents the evolution of the upper bound. This bound is the cost of the current incumbent (though not necessarily feasible) solution obtained either as a result of sampling the quantum state resulting from the VQA execution at that node or as a result of variable fixing.
The red line with circular markers represents the evolution of the lower bound. 
We see that both upper and lower bounds converge as the number of nodes explored increases. For this particular instance, an optimal solution was found after exploring $18$ nodes. Convergence is guaranteed by design, though we provide more explanation of that in Appendix \ref{app:convergence}. 
Here, on Figure \ref{fig:gap}, we do not use the actual time on the horizontal axis. It is possible, but less informative at the moment as we can only access simulators. Measuring the real primal-dual integral, considering the objective value of a function of time, would be a helpful measure on real hardware for comparing different algorithms on the same hardware. 

\begin{figure}[h]
    \centering
    \includegraphics{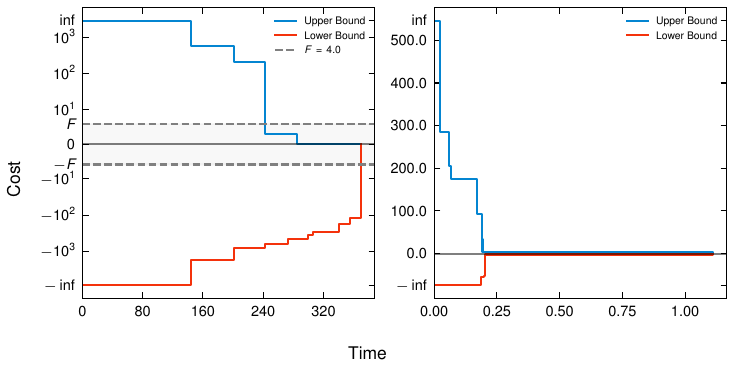}
    \caption{Comparison with SCIP. The blue line shows the upper bound (cost of incumbent solution) as a function of time in seconds. The red line shows the lower bound as a function of time. A solid black line marks the optimum, aligned to $0$, with dashed black lines representing the cost $F$ of the worst feasible solution, in this case $4$. Between the dashed lines, the y-axis is scaled linearly, while outside the dashed line, a logarithmic scale is used.}
    \label{fig:scip-time}
\end{figure}

On the other side, the calculation as a function of processed nodes hides an important property of the primal-dual integral. For an exact, time-based comparison between classical branch-and-bound based solvers and our algorithm, we conducted further experiments. For them, we used the classical solver SCIP \cite{Achterberg2009, SCIPOptSuite10}, with no presolve, heuristics, cuts, and propagation for fairness. On Figure \ref{fig:scip-time}, we show the primal-dual integral of the QCBB solver run on an instance with 18 variables.

With the x-axis now representing time, we can clearly see a benefit to solving smaller and smaller problem instances as the solver progresses toward the optimal solution. The time between node evaluations shrinks, since as the problem itself shrinks, the computational cost of running (in this case, simulating) the quantum circuit also decreases. 

One might notice the difference in the total runtimes. It is indeed a huge difference; however, a direct comparison is still unfair, since the quantum resources we can use are simulated with limited capacity, and the instances are not the same. The plot is for demonstration purposes at this point, it represents one of our goals: the algorithms are comparable with the same performance metrics. As quantum hardware develops and the algorithms within the hybrid branch-and-bound framework mature, this measure could become increasingly relevant.

\begin{figure}[h]
    \centering
    \includegraphics{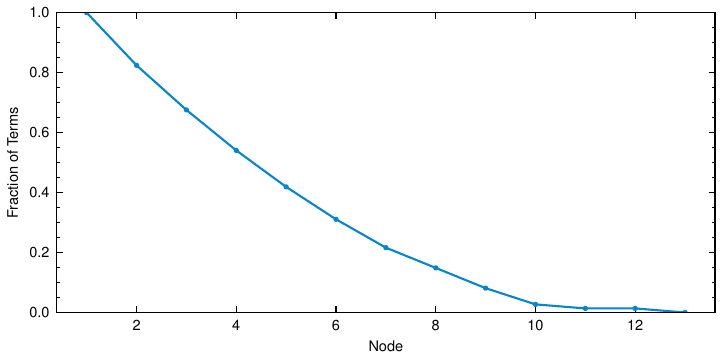}
    \caption{Fraction of many-body terms. A blue line with circular markers shows the fraction of many-body terms remaining (compared to \eqref{eq:MH}) as a function of nodes evaluated.}
    \label{fig:terms}
\end{figure}

As explained in Section \ref{sec:branching}, the subproblem in any node has at least one variable fixed (potentially more due to further reductions), which not only results in a subproblem with fewer variables, but also one with fewer many-body interactions which can greatly impact the size of the quantum circuit not only in terms of the number of qubits required, but also the number of costly two-qubit gates.
We show, in figure \ref{fig:terms}, the fraction of many-body terms present, compared to \eqref{eq:MH}, as the tree is being explored as a blue line with circular markers.
Since the plot does not preserve the tree structure, we can expect to see occasional increases in the fraction of terms remaining at a given node.
The number of many-body terms at a node is always less than that in the parent node, but is not overall expected to decrease monotonically as other parts of the tree with more many-body terms might be explored during the optimization process.

To demonstrate the value of solving reduced subproblems, we can look at how our algorithm compares to just running QAOA in terms of the reducing cost associated with the quantum state reached after optimization. 
To compare different nodes in the tree, we must evaluate the result in the context of the root node (either \eqref{orig} or \eqref{eq:MH}), which we do by adding back constant terms arising from reductions as explained in Section~\ref{sec:subproblem}.
Since we use QAOA within our branch and bound approach, restricting the algorithm to evaluating only a single (the root) node would be equivalent to simply relying on QAOA to solve the problem. 
In our experiments, we utilize the COBYLA optimizer with stricter convergence criteria, which would result in more iterations and consequently more queries being made to a quantum computer before termination of the algorithm.
This is to show that splitting up these iterations over many nodes and subproblems is a better use of quantum computational resources than just expending a similar amount of compute on just the master problem.
To this end, we configure the optimizer with a maximum number of $50$ iterations per node for our branch and bound algorithm and at most $500$ iterations for plain QAOA.
The cost values are adjusted such that the optimum cost is $0$ and the distance between the optimum and the worst feasible solution is $1$.

\begin{figure}[h]
    \includegraphics{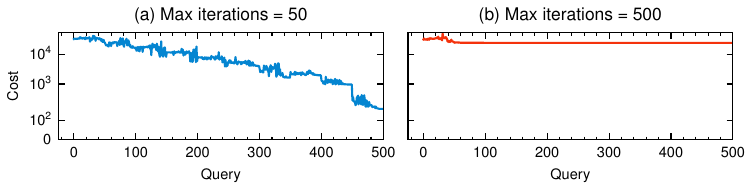}
    \caption{Comparison of branch and bound with plain QAOA. The plots show the expected cost associated with the quantum state, as a function of queries made to a quantum simulator, as it evolves during the execution of the VQA in each node. Subplot (a), in blue, uses the branch and bound approach proposed in this work with a limit of $50$ iterations per node. In subfigure (b), in red, only QAOA is used. The cost at the optimum is adjusted to $0$.}
    \label{fig:expval}
\end{figure}

In our testing, we see a clear advantage attributable to the subproblem reduction step as the expected cost associated with the quantum state drops as we switch to solving smaller and smaller subproblems in Subfigure (a) of Figure \ref{fig:expval}. In contrast, with plain QAOA (Figure \ref{fig:expval} (b)), the improvements plateau out fairly early and at a higher expected cost, which can suggest a high probability of sampling suboptimal variable assignments.
It is clear to see the necessity of appropriately configuring the optimizer used, so that we start branching before the improvements plateau out.
Notably, the subproblem typically begins with a higher expected cost than that reached in the parent node, but tends to end up with a lower expected cost by the end of the optimization process. This opens up the question of whether parent nodes can inform the starting parametrization of child nodes.

\section{Discussion}
\label{sec:disc}

We introduced the framework for a hybrid quantum branch and bound algorithm. Since there are many parts to it, also from different fields, we aimed for simplicity and understandability. With that in mind, we would like to have a slightly more elaborate discussion section to address many different characteristic features and the possible improvements of the algorithm.

\subsection{Classical algorithmic opportunities}

Conflict values have a similar structure as reduced costs have in linear programming. While in linear programming, reduced costs drive towards optimality, in this algorithm, conflict values drive towards feasibility. With this value, we aim to reduce the complexity of the problem at each level of the tree as much as possible. Furthermore, this is a reduction that inherently takes differences in the chosen quantum subroutines into account. As they might find different properties of the energy landscape harder to address, the conflict values aim to capture this hardness and resolve it by removing the right variable and branching based on its domain.

There are improvement opportunities inside the algorithm: the value of $M$ does not change going down the tree. However, with each step, one could restrict and reduce $M$ more and more, and consequently get a problem which is numerically better, the extremes are smaller, and both the VQA and the GW-bound are better. One caveat is, computing the global bound requires computing and maintaining the change of $M$ over the nodes, for example, according to \cite{Alessandroni2025}, just like we track the change of the constant offset of $H$. Some of the further opportunities are improving the node selection \cite{Hansen1990}, adding more sophisticated subproblem feasibility testing \cite{Savelsbergh1994, Boros2002}, adding proper presolving, and improving the approximate bound calculation by looking at graph structure or using quadratic programming relaxations to the QUBO at hand \cite{Rehfeldt2023, Punnen2022}.

\subsection{Quantum algorithmic opportunities}

The algorithm has potential for further development and large improvements on the quantum side as well. One could use more sophisticated VQAs \cite{Cheng2024, Zhang2024}, which could significantly enhance the overall performance, as it is the heart of the algorithm. One could look for any suitable algorithm from the literature and measure its impact. 

One step closer to the quantum part itself, one could find that the problem encoding we use is naive and simple. Another large improvement can be achieved by better problem encoding and the use of quantum resources \cite{Weidenfeller2022, MontaezBarrera2024, Chatterjee2024, Kanatbekova2025, Bak2025}.

Taking one more step, and looking into further details, one could also see that they could use more clever transpilation and compilation methods \cite{TuanHai2024} and circuit optimization techniques \cite{Zhu2024}. That we also did not consider here.

\subsection{Algorithm characteristics}

There are also some aspects of usability that we would like to discuss. First, while BLPs are NP-complete \cite{Karp1972}, and thus in theory any MILP can be reduced to it in polynomial time. Practically, our required problem structure is strict and tight, however, reductions to it can be useful, and this structure can be exploited when formulated well. 

One caveat is that one needs to train a VQA at each node. In general, that is expensive and slow. However, we do not rely strictly on their solution quality. One might exploit this fact by using shallow circuits and setting a hard limit on training iterations. That would entail worse quality solutions on the subproblems and thus more possible nodes to evaluate. However, since the subproblems get smaller and simpler, even shallow VQAs could solve them.

Since we build the tree of variable assignments, the algorithm guarantees optimality, even though not in polynomial time, as the number of subproblems can be exponential. On the other hand, by solving the full problem at each node and reducing complexity at each level, our algorithm might find good quality solutions very early. This makes it an exciting candidate as a hybrid advice for classical solvers. Another advantage of this behavior is making quick decisions of a good enough quality, which is of great interest in industrial optimization.

In Table \ref{tab:opportunities}, we give an overview of the current solutions components and list the possible opportunities for improving the method.

And finally, even though our algorithm has many possibilities for improvement, our results prove that even in this early state, it can produce measurable, good-quality results with explainable, mathematical guarantees.

\begin{table}[t]
\setlength{\extrarowheight}{2pt}
\centering
\caption{Overview of current solution components and opportunities.}
\label{tab:opportunities}
\begin{tabularx}{\textwidth}{l X X}
\hline
\textbf{Algorithmic part} & \textbf{Current solution} & \textbf{Opportunities} \\[2pt]
\hline

Hamiltonian encoding
& Via QUBO penalty terms
& Qubit-efficient encoding \cite{Kanatbekova2025},\\
& & Improved constraint encoding for inequalities \cite{MontaezBarrera2024},\\
& & Logarithmic-space encoding \cite{Chatterjee2024} \\[6pt]

Quantum optimization
& QAOA
& LR-QAOA \cite{MontaezBarrera2025}, \\
& & D-Wave annealers,\\
& & KIPU DCQO-based approaches \cite{Romero2025},\\
& & varQITE-based approaches \cite{Motta2019, Soley2021, Gacon2024} \\[6pt]

Branching
& Most conflicting variable
& Least conflicting variable for better propagation,\\
& & Inference branching \cite{Achterberg2009}, \\
& & VSIDS branching \cite{Moskewicz2001} \\ [6pt]

Problem reduction
& Variable bound propagation
& SAT-based techniques \cite{Moskewicz2001, Aloul2006} \\[6pt]

Node selection
& Lowest bound
& Pseudo-boolean node scores \cite{Hansen1990},  \\
& & DFS-based selection\\[6pt]

Bounding algorithm
& Goemans--Williamson approx.
& More sophisticated QUBO-based algorithms with bound guarantees \cite{Punnen2022, Dunning2018},\\
& & Efficient algorithms for MaxCut \cite{Rehfeldt2023} \\[6pt]

Infeasibility checking
& $M$-based proof
& BLP-based proofs \cite{Savelsbergh1994, Boros2002} \\[2pt]
\hline
\end{tabularx}
\end{table}


\subsection{Implementation and integration}

As our resources have been greatly limited, we are not able to provide extensive testing on real hardware, nor on scaled-up problems. That is, we look forward to tackling problems from QOBLIB \cite{QOBLIB} with our approach with appropriate computational capabilities. We provided our results as numerical proofs, and provided the code as part of an open-source project at \url{https://github.com/david-sipos/qcbb}.

\backmatter



\bibliography{main}

\onecolumn\newpage
\appendix

\section{Binary Linear Program to Ising Hamiltonian}

We show the calculations of how we transform the binary linear program to Ising Hamiltonian. This is a very common conversion, however, we use some of the elements in the main text, therefore, we show this derivation explicitly. 

\subsection{Conversion}
\label{app:BLPtoH}

As a first step, we create a QUBO form from \eqref{orig} by moving the constraints to the objective as
\begin{align}
    Ax=b ~\longmapsto& ~\quad (Ax-b)^T(Ax-b) \\ 
    &= (x^TA^TAx - 2(Ax)^Tb + b^Tb) \\ 
    &= x^T(A^TA)x - 2(Ax)^Tb + b^Tb
\end{align}
so the problem becomes, with a sufficiently large penalty factor, $M$
\begin{equation}
    \min~ \left\{x^T(MA^TA)x + c^Tx - 2M(Ax)^Tb + Mb^Tb ~\bigm|~ x\in\left\{0,1\right\}^n\right\}.
\end{equation}
Now we change the domain from $x\in\{0,1\}$ to $\sigma\in\{-1,1\}$ by substituting $x = \frac{\sigma + \1}2$, 

\begin{align}
    H({x}) =~ & x^T(MA^TA)x + (c^T - 2Mb^TA)x + Mb^Tb \\
    =~ & \frac{\sigma^T + \1^T}2(MA^TA)\frac{\sigma + \1}2 + (c^T - 2Mb^TA)\frac{\sigma + \1}2 + Mb^Tb \\
    =~ & \frac{1}{4} M \sigma^T (A^T A) \sigma 
    + \frac{1}{2} M \mathbf{1}^T (A^T A) \sigma 
    + \frac{1}{4} M \mathbf{1}^T (A^T A) \mathbf{1} \\
    & ~+ \frac{1}{2}  c^T\sigma + \frac{1}{2} \mathbf{1}^T c
    - M  b^T A \sigma - M b^T A \mathbf{1}
    + M b^T b \\
    =~ & \frac{1}{4} M \sigma^T (A^T A) \sigma + \frac{1}{2} \left(c^T - 2 Mb^T A + M\mathbf{1}^T (A^T A) \right) \sigma \\
    & ~+ \frac{1}{4} M \mathbf{1}^T (A^T A) \mathbf{1} + \frac{1}{2} \mathbf{1}^T c - M b^T A \mathbf{1}
    + M b^T b
\end{align}
and then the problem resolves to
\begin{equation}
    H(\sigma) = - \sum_{i<j}J_{ij}\sigma_i \sigma_j - \mu \sum_j h_i \sigma_i + C,
\end{equation}
where
\begin{equation}
    J = -\frac{1}{4}MA^TA
\end{equation}
and
\begin{equation}
    h = c^T - 2 Mb^T A + M\mathbf{1}^T (A^T A)
\end{equation}
with
\begin{equation}
    \mu = -\frac{1}{2}.
\end{equation}
The constant term $C$ does not affect the procedure and thus can be put aside during optimization and add back after. The constant is
\begin{equation}
    C = \frac{1}{4} M \mathbf{1}^T (A^T A) \mathbf{1} + \frac{1}{2} \mathbf{1}^T c - M b^T A \mathbf{1}
    + M b^T b.
\end{equation}

Here we arrived at an Ising Hamiltonian $H(\sigma)$ with $\sigma$ noting its spin configuration.

\subsection{Big M value}
\label{app:bigM}

Big $M$ should be a sufficiently large constant. For any feasible $x$ and any infeasible $x'$,
\begin{equation}
    c^Tx < c^Tx' + M\left(Ax'-b\right)^T\left(Ax'-b\right).
\end{equation}
must hold. 
In \cite{Czegel2025} we have derived a big $M$ value, for the numerical precision $\kappa$ needed for $A$ and $b$,
\begin{equation}
    M = \frac1{\kappa}\sum_i |c_i| \geq 
    \frac{\max_{x, x'} c^T\left(x-x'\right)}{\min_{x'}\left(Ax'-b\right)^T\left(Ax'-b\right)} 
    \geq \max_{x, x'} \frac{c^T\left(x-x'\right)}{\left(Ax'-b\right)^T\left(Ax'-b\right)}. \label{eq:bigM}
\end{equation}

\section{Bounding calculations}
\label{app:bound}

Here, we describe the exact calculation behind the bounding step. We calculate a lower bound using an approximate solution to the Maximum Cut problem.

\subsection{Maximum Cut problem}

 This problem is defined on an undirected graph $G = (V, E)$ with weights on its edges $w_{uv} \in\mathbb{R} \setminus \{0\}, (u,v) \in E, u,v\in V$. We denote edges between sets $U,W\subset V$ by $(u,v) \in E(U,W)$, where $u\in U, v\in W$ and $(u,v) \in E$. Now, a cut $g(S), S\subseteq V$ is a bipartition of the vertices, where the value of the cut is given by the sum of weights between the partitions:

\begin{equation}
    g(S) = \sum_{
    (u,v)\in E(S,V\setminus S)} w_{uv}.
\end{equation}
The problem is to find the cut with the greatest value, i.e.
\begin{equation}
    z^* = \max_{S \subseteq V} g(S) = \max_{S \subseteq V} \sum_{(u,v)\in E(S,V\setminus S)
    } w_{uv}.
\end{equation}

\subsection{Reduction of $H(\sigma)$ to Maximum Cut}
\label{app:reduction}

Recall that $H$ is an Ising Hamiltonian operator over $n$ spins $\sigma \in \{1, -1\}^n$, and has the form of
\begin{equation}
\label{eq:ham_gen}
    H(\sigma) = - \sum_{i<j}J_{ij}\sigma_i \sigma_j - \mu \sum_j h_i \sigma_i.
\end{equation}
We derive the following reduction to the Maximum Cut problem, based on the reduction from \cite{Barahona1989}. Given a Hamiltonian with $J$ and $h$ as on \eqref{eq:ham_gen}, we define the following graph: $V = \{0, 1, \dots, n\}$, $E = \left\{(0,i) ~|~ i\in V\setminus\{0\}, h_{i} \neq 0\right\} \cup \{(i,j) ~|~ i,j\in V\setminus\{0\}, J_{ij} \neq 0, i< j\}$ with edge weights $w_{ij} = J_{ij} ~ \forall (i,j)\in E, i\neq 0$, and $w_{0i} = \mu h_i ~ \forall i\in V\setminus\{0\}$, where $h_i \neq 0$. 

The domain of the variables is still $\{-1,1\}$. Interpreting it in the domain of the binary Maximum Cut problem, for each assignment of the variables $\sigma_i$, let us organize them to two disjoint sets of vertices of the graph. Since the problem is symmetric, we can choose vertex 0 to be part of any of the sets. Then, defining $V^+ = \{i \in V \mid \sigma_i = 1\} \cup \{0\}$ and $V^- = \{i \in V \mid \sigma_i = -1\}$, by construction we get the following, equivalent minimization problem:
\begin{equation}
 \min H(\sigma) = \min \sum_{(i,j) \in E(V^+, V^+)} w_{ij} + \sum_{(i,j) \in E(V^-, V^-)} w_{ij} - \sum_{(i,j) \in E(V^+, V^-)} w_{ij}
\end{equation}
which is equivalent to
\begin{equation}
    \min \sum_{(i,j) \in E}w_{ij} - 2 \sum_{(i,j) \in E(V^+, V^-)} w_{ij}.
\end{equation}
Now, if we take a look at $\sum_{(i,j) \in E(V^+, V^-)} w_{ij}$, this is exactly the objective value from the formulation of the Maximum Cut, where $V^+ = S$ and which we denoted above by $g(V^+)$. If we take the optimal solution, $z^* = \max_{V^+ \subseteq V} g(V^+)$, we arrive at
\begin{equation}
\label{eq:reduction_result}
    \min H(\sigma) = -2 z^* + W,
\end{equation}
where
\begin{align}
    W = \sum_{(i,j) \in E}w_{ij} = \sum_{i < j} J_{ij} + \mu\sum_i h_i.\label{eq:W-J-h}
\end{align}

\subsection{Approximate solution to Maximum Cut}

The Maximum Cut problem is NP-hard \cite{garey1979}, its decision version is NP-complete \cite{Karp1972}. It is also APX-hard \cite{Papadimitriou1991}, meaning there is no efficient $\epsilon$-approximation algorithm exists unless P=NP. However, there is an efficient, semidefinite programming (SDP) relaxation-based randomized algorithm by Goemans and Williamson \cite{Goemans1995} with a worst-case average approximation ratio $\alpha = 0.87856$. That means, for an undirected, positively weighted graph $G$, with maximal cut value $z^*$ and the cut value from the Goemans-Williamson algorithm $z_{GW}$, we get

\begin{equation}
    \alpha z^* \leq z_{GW} \leq z^*,
    \label{eq:gwboundorig}
\end{equation}
 so that we can use $z_{GW}$ for bound calculation.
\subsection{The lower bound}

Here, we provide the bound calculation as well as a lower bound to the bound itself. 
Since we have negative edge weights as well in the graph, we need the following Theorem 3.2.1 from \cite{Goemans1995}. For the graph $G = (V, E)$, encoding the problem, we have defined $W$ as the sum of the edge weights, 

\begin{equation}
\label{eq:W}
    W = \sum_{(i,j) \in E}  w_{ij} ,
\end{equation}
We now extend that definition to graphs with negative edge weights by defining
the total negative edge weights, 
\begin{equation}
\label{eq:W-}
    W^- = \sum_{(i,j)\in E : w_{ij} < 0} w_{ij},
\end{equation}
and we will use the already introduced notation of $\alpha = 0.87856$ for the average approximation ratio of the Goemans-Williamson algorithm. In this section, we derive a lower bound to $\min H(\sigma)$, of the form of \eqref{eq:MH} (and \eqref{eq:RMH}), using these notations.

\begin{theorem}[\cite{Goemans1995}, Theorem 3.2.1.]
\label{thm:negweight}
    Let $z_{GW}$ the expected value of the maximum cut on $G$ from the algorithm, and $z^*$ the maximal cut value, then
    $$z_{GW} - W^- \geq \alpha \left(z^* - W^-\right).$$ 
\end{theorem}

\begin{theorem}
    Given an Ising Hamiltonian subproblem $\min H(\sigma)$ of the form of \eqref{eq:MH}, and its reduction to Maximum Cut on graph G. Let $z_{GW}$ denote the solution from the Goemans-Williamson algorithm on G. Then, 
    \[LB_{GW}(H(\sigma)) = -\frac{2}{\alpha} z_{GW} + \left(\frac{2}{\alpha}-2\right) W^- + W\]
    is a lower bound to $\min H(\sigma)$, 
    i.e.$$ LB_{GW}(H(\sigma)) \leq \min H(\sigma).$$
\end{theorem}

\begin{proof}
    According to Theorem \ref{thm:negweight}, in our case, where negative edge weights are also considered, we can derive that
    \begin{equation}
    \label{eq:approxR}
        z^* \leq \frac{1}{\alpha} z_{GW} - \left(\frac{1}{\alpha} - 1\right) W^-.
    \end{equation}

    Let us consider the reduction in Section \ref{app:reduction} from an Ising Hamiltonian $H$ to a Maximum Cut problem on graph $G$. Using the Goemans-Williamson algorithm to find the maximum cut on $G$ and by combining \eqref{eq:reduction_result} and \eqref{eq:approxR}, we get
    \begin{align}
        \min H(\sigma) &= -2z^* + W \\
        &\geq -2 \left(\frac{1}{\alpha} z_{GW} - \left(\frac{1}{\alpha} - 1\right) W^-\right) + W \\
        &= -\frac{2}{\alpha} z_{GW} + \left(\frac{2}{\alpha}-2\right) W^- + W 
    \end{align}
    Thus, a valid lower bound is
    \begin{equation}
    LB_{GW}(H(\sigma)) = -\frac{2}{\alpha} z_{GW} + \left(\frac{2}{\alpha}-2\right) W^- + W.  \label{eq:H_GW}
    \end{equation}
\end{proof}

\begin{theorem}\label{thm:LB}
    The lower bound $LB_{GW}(H(\sigma))$ cannot be worse then 
    $$LB = \frac1\alpha \min H(\sigma) - \left(\frac{1-\alpha}{\alpha}\right) \left(W-2W^-\right)$$
    calculated from the optimum, that is, $$LB\leq LB_{GW}(H(\sigma)) \leq \min H(\sigma).$$
\end{theorem}
    
\begin{proof}
    
    The error of the approximation, i.e.\ the difference between the optimal solution value $\min H(\sigma)$ and $LB_{GW}(H(\sigma))$ is given by
    \begin{align}
    \Delta_{approx} &= \min H(\sigma)-LB_{GW}(H(\sigma))  \\
        &= -2z^* + W - \left(-\frac{2}{\alpha} z_{GW} + \left(\frac{2}{\alpha}-2\right) W^- + W\right)  \\
        &=-2z^* + \frac{2}{\alpha} z_{GW} - \left(\frac{2}{\alpha}-2\right) W^- . 
    \end{align}
    Now this shows an interesting behavior. Since $z_{GW} \leq z^*$, given any $H$, the largest difference from the optimal solution value so the worst bound occurs when $z_{GW} = z^*$. The better the result the approximation algorithm produces, the worse the bound we obtain:
    \begin{align}
       \max \Delta_{approx} &= \max_{z_{GW}} \left(-2z^* + \frac2\alpha z_{GW} - \left(\frac2\alpha - 2\right) W^-  \right) \\[6pt] &=  \left(\left(\frac2\alpha - 2\right) z^* - \left(\frac2\alpha - 2\right) W^-  \right) 
        = \left(\frac2\alpha - 2\right) \left(z^* - W^-\right).
        \label{eq:worstbound}
    \end{align}
    Therefore the bound $LB_{GW}(H(\sigma))$ is in the worst case, due to \eqref{eq:H_GW} and \eqref{eq:worstbound}, 
    
    \begin{equation}
       \min H(\sigma) - \left(\frac2\alpha - 2\right) \left(z^* - W^-\right) \leq LB_{GW}(H(\sigma)) \leq  \min H(\sigma)
    \end{equation}
    The worst bound, in terms of the elements of the Ising Hamiltonian $H$ is, since $z^* = -\frac12\bigg(\Big(\min H(\sigma)\Big) - W\bigg)$,
    \begin{align}
        LB &= \min H(\sigma) - \left(\frac2\alpha - 2\right) \left(z^* - W^-\right) \\
        &= \min H(\sigma) + \frac12 \left(\frac2\alpha - 2\right) \left(\min H(\sigma) - W + 2W^-\right) \\
        &= \frac1\alpha\min H(\sigma) + \left(\frac1\alpha - 1\right) \left( - W + 2W^-\right),
    \end{align}
    what had to be shown. 
\end{proof}

\begin{remark}
    The result of Theorem \ref{thm:LB} can be written in terms of the Hamiltonian, i.e.\ 
    \begin{equation}
    LB = \frac1\alpha \min H(\sigma) - \left(\frac1\alpha - 1\right) \left(\sum_{i<j}|J_{ij}| + \mu\sum_i |h_i|\right).
    \end{equation}
    This is due to the construction of the graph, and its weights $W$ \eqref{eq:W} and $W^-$ \eqref{eq:W-}. 
    
    As $W$ is the sum of all weights while $W^-$ is the sum of negative weights, it is easy to check that $W-2W^-$ gives the sum of the weights in absolute values. In formula, following \eqref{eq:W-J-h} 
    \[
     W-2W^- = \sum_{(i,j)\in E} w_{ij} -2 \sum_{\substack{(i,j)\in E\\ w_{ij}<0}} w_{ij} = \sum_{(i,j)\in E} |w_{ij}| = \sum_{i<j}|J_{ij}| + \mu\sum_i |h_i|.
    \]
        
\end{remark}

\section{Infeasibility proof using M}
\label{app:infeas}

Based on the bound calculation above, we can derive a condition that is sufficient to prove a subproblem infeasible.

\begin{statement}
    When $LB_{GW}(\hat H) + \hat C \geq M$, the subproblem, encoded in $\hat H$, is infeasible.
\end{statement}
\begin{proof}
    It is by the design of the calculations of $M$ and the lower bound. 
    
    By the construction in \eqref{eq:bigM}, $M$ dominates all feasible solution values, meaning that for each feasible solution $\hat x$, and its value $\hat c^T \hat x = \hat H + \hat C$, it is true that 
    \begin{equation}
        \hat c^T \hat x < M. \label{eq:cxM}    
    \end{equation}
    
    As for the lower bound, $LB_{GW}(\hat H)$ is a lower bound  to $\min_{\hat x} \hat{c}^T \hat x - \hat C = \min_{\hat \sigma} \hat H$. That means, for any feasible solution $\hat x$, it holds that 
    \begin{equation}
        \hat c^T \hat x \geq LB_{GW}(\hat H) + \hat C. \label{eq:LB_M}
    \end{equation}

    Let us suppose that there exists a feasible solution $\hat x$ for which $LB_{GW}(\hat H) + \hat C \geq M$. Then by \eqref{eq:bigM} and \eqref{eq:LB_M} we have
    \begin{equation}
        LB_{GW}(\hat H) + \hat C \geq M > \hat c^T \hat x \geq LB_{GW}(\hat H) + \hat C.
    \end{equation}
    Combining the two gives a contradiction, therefore, there is no feasible solution for the subproblem if its lower bound is at least $M$.
\end{proof}

\section{Number of many-body terms over the levels}
\label{app:manybody}

\begin{statement}
    The number of many-body terms does not increase at each sub-problem creation step. Formally, following the notation of Section \ref{sec:branching}, the number of non-zero elements above the main diagonal in the child node $\tilde J$ is less than or equal to their number in the parent node $\hat J$.
\end{statement}

\begin{proof}
    Since $J = -\frac{1}{4}MA^TA$, we need to take a look at $\hat{A}$ and $\Tilde{A}$. We get a zero in $A^TA$ if either two columns of $A$ have a zero inner product or at least one of the columns is a zero vector. We choose variables that have the greatest conflict values. We can exclude the latter, all-zero columns, since by construction, we would never choose a variable to fix that does not appear in any constraints.

    For the pairwise orthogonal columns, we can only guarantee that their number does not decrease by more than $\hat{n}-1$. Let us suppose that we remove column $k$ from $\hat{A}$. If that column had zero inner product with each of the other columns, we would end up removing $\hat{n}-1$ zeros from above the main diagonal of  $\hat{A}$ when constructing $\Tilde{A}$. Since the difference between the above-diagonal elements of $\hat{J}$ and $\Tilde{J}$ is exactly this set of entries, we did not remove any non-zeros from $\hat{J}$. 

    Any other cases would include removing non-zeros from above the main diagonal, and thus removing many-body terms from the operator. That is, because the removal operation does not affect any other elements but the ones in the column $k$, which we remove. 
\end{proof}

In the latter case, the variables are completely independent of each other. That case could, on one hand, slow our solution process down. On the other hand, it is an opportunity for decomposition of the problem.

\section{Convergence of the bounds}
\label{app:convergence}

Here we show that both the solutions that we find and the lower bound, calculated by the Goemans-Williamson algorithm, converge to the optimal solution value of \eqref{orig}. This is a criterion to use the primal-dual integral \cite{Berthold2013} to measure the solution process quality, and also to provide a comparison to classical solvers.

The statement is a consequence of the construction of the algorithm. At any point, the frontier of the tree includes both the current best solution (incumbent, upper bound) and the current global lower bound. We also know the exact relation between the bounds, as the incumbent must be greater than or equal to the lower bound, see Appendix \ref{app:bound}. 

Since at each subproblem creation, we fix at least one variable and remove it from further optimization, the bound calculation becomes more precise. At the same time, the lower bound itself is also at least equal to the lower bound of the problem in the parent node by definition in \eqref{eq:LB}, and thus in the root. Formally, using the notation of Section \ref{sec:subproblem},

\begin{equation}
    LB(\hat{H}) + \hat{C} \leq LB(\Tilde{H}) + \Tilde{C}.
\end{equation}
When there are no more free variables left in the subproblem, either the two local bounds meet, or we prove the subproblem infeasible. If the bounds meet, that is formally due to there is nothing to approximate to get a lower bound, and thus it becomes $0$ + fixed values. That is exactly the local upper bound. On a side note, we can end up pruning the sub-tree completely, which does not affect global convergence since we do not lose optimal solutions by pruning the sub-tree.

Putting the pieces together, on each path from the root node, the complete problem, to a leaf node, a subproblem with no variables, we either have a proof of infeasibility, non-improvement, or a proof of local optimum.

We can evaluate all the nodes on all such paths from the root to any of the end nodes, infeasible, pruned, or leaf. Due to the removal of at least one variable on each level, the number of nodes to evaluate is $2^n-1$. In the end, we can enumerate all the leaf nodes and look for the best solution, as that will provide a global optimum.

Of course, computationally this does not help, but proves that for each problem we can obtain either a proof of infeasibility, a proof of an existing better solution, or a proof of optimality.

\section{Further numerical results and algorithm characteristics}
\label{app:results}

Here we show results obtained on a variety of problem instances, similar to those shown in Section \ref{sec:results}.
We've opted to showcase a range of problem instances that range from easier to more difficult, as judged by the number of node evaluations required for the bounds to converge.

\begin{figure}[h]
    \centering
    \includegraphics{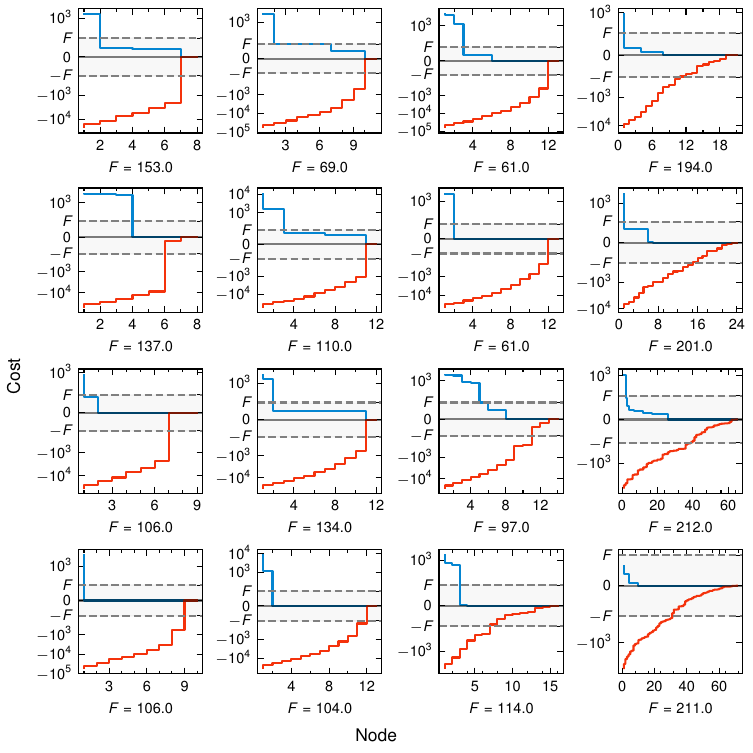}
     \caption{Convergence of bounds for $16$ problem instances. Similar to Figure \ref{fig:gap}, the blue line shows the upper bound (cost of incumbent solution) as a function of the number of nodes evaluated. The red line shows the lower bound as a function of the number of nodes evaluated. A solid black line marks the optimum, aligned to $0$, with dashed black lines representing the cost $F$ of the worst feasible solution, with its value shown under each subfigure. Between the dashed lines, the y-axis is scaled linearly, while outside the dashed line a logarithmic scale is used.}
     \label{fig:multi-gap}
\end{figure}

As in Figure \ref{fig:gap}, Figure \ref{fig:multi-gap} showcases the convergence of the upper bound, given by the incumbent solution at a given node, and the lower bound, calculated using the approximate solution of the corresponding MaxCut instance.
Possible use of these bounds is as a termination criterion or as a quality measure of the process of the solver. In our testing, it seems that for easier-to-solve instances, the gap closes suddenly as the optimal solution is found, likely as a result of variable propagation. In more difficult instances, we see the gap closing in sooner than the last couple of iterations. For such problems, the node-selection and variable-selection processes can greatly impact the performance of the algorithm.

\begin{figure}[h]
    \centering
    \includegraphics{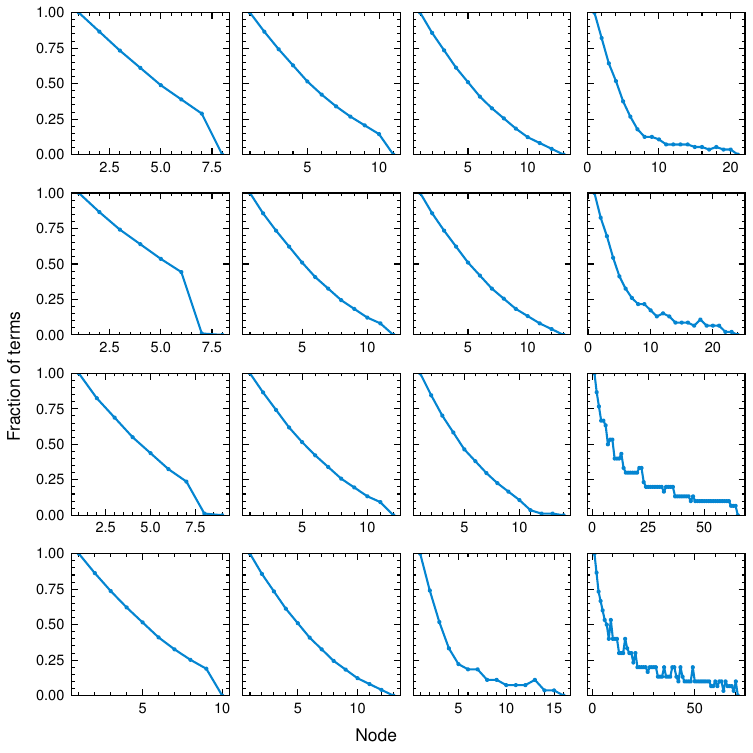}
    \caption{Fraction of many-body terms for $16$ problem instances. Similar to Figure \ref{fig:terms}, a blue line with circular markers shows the fraction of many-body terms remaining (compared to \eqref{eq:MH}) as a function of nodes evaluated.}
    \label{fig:multi-terms}
\end{figure}

Similar to Figure \ref{fig:terms}, in Figure \ref{fig:multi-terms} we show the fraction of many-body terms present, compared to \eqref{eq:MH}, as the tree is being explored as a blue line with circular markers.
In some instances, we can see that due to the order of evaluations, and thus the plot itself, not following the tree structure, occasional increases are present as the algorithm explores nodes higher up in the tree or those where variable propagation resulted in fewer variables being removed.

\begin{figure}[h]
    \centering
    \includegraphics{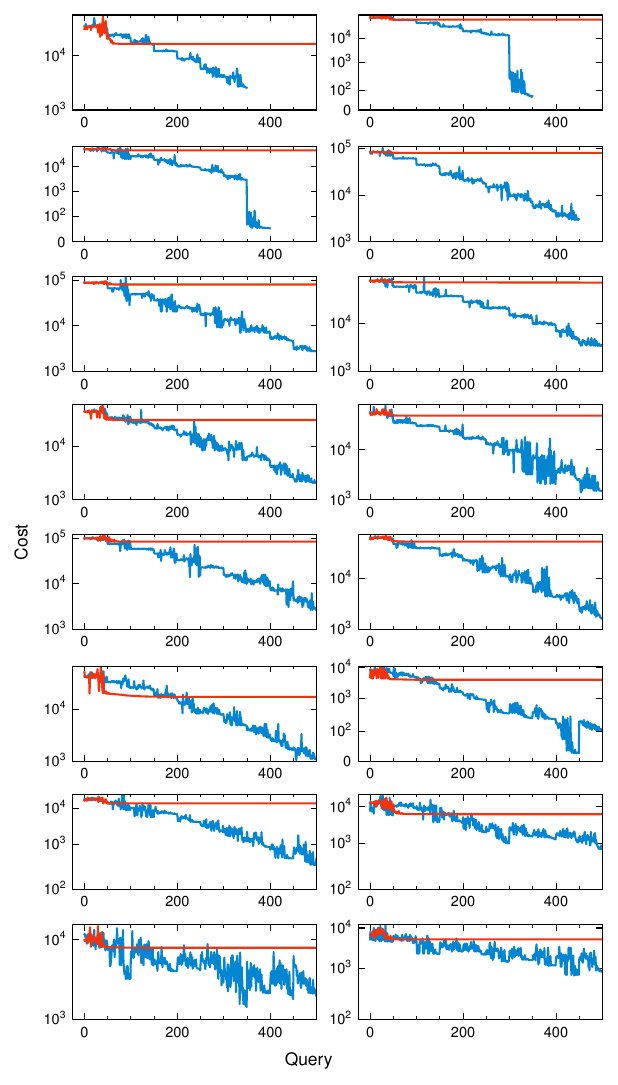} 
    \caption{Comparison of branch and bound with plain QAOA for $16$ problem instances. The plots show the expected cost associated with the quantum state, as a function of queries made to a quantum simulator, as it evolves during the execution of the VQA in each node. Values on the y-axis show distance from the optimum. For each instance, plotted in blue, we use the branch and bound approach proposed in this work with a limit of $50$ iterations per node. For the red line, only QAOA is used.}
    \label{fig:multi-cost}
\end{figure}

As with the instance in Figure \ref{fig:expval}, in Figure \ref{fig:multi-cost} we continue to see a clear advantage attributable to the subproblem reduction step as the expected cost associated with the quantum state drops (shown in blue) as we switch to solving smaller and smaller subproblems. With plain QAOA, the improvements plateau out fairly early, and while it can achieve lower expected cost with fewer queries, in the long run, our approach achieves lower expected cost, which can suggest a high probability of sampling favorable variable assignments.
Here, we can still observe the potential impact the specifics of the VQA used can have. In many cases, we see only a slow improvement attributable to the VQA optimization process, with larger intermittent drops likely the effect of subproblem reductions. Note that expected costs can be greatly skewed by sampling infeasible solutions, which have very high costs due to penalty terms introduced as part of the QUBO formulation.

\end{document}